# New paper-by-paper classification for Scopus based on references reclassified by the origin of the papers citing them

Jesús M. Álvarez-Llorente[a], Vicente P. Guerrero-Bote[b], Félix de Moya-Anegón[c]

[a] Department of Computer and Telematic Systems Engineering, University of Extremadura, Badajoz, Spain, llorente@unex.es

[b] Department of Information and Communication, University of Extremadura, Badajoz, Spain, guerrero@unex.es

[c] SCImago Research Group, Granada, Spain, felix.moya@scimago.es

**Abstract**

A reference-based classification system for individual Scopus publications is presented which takes into account the categories of the papers citing those references instead of the journals in which those cited papers are published. It supports multiple assignments of up to 5 categories within the Scopus ASJC structure, but eliminates the Multidisciplinary Area and the miscellaneous categories, and it allows for the reclassification of a greater number of publications (potentially 100%) than traditional reference-based systems. Twelve variants of the system were obtained by adjusting different parameters, which were applied to the more than 3.2 million citable papers from the active Scientific Journals in 2020 indexed in Scopus. The results were analyzed and compared with other classification systems such as the original journal-based Scopus ASJC, the 2-generation-reference based M3-AWC-0.8 (Álvarez-Llorente et al., 2024), and the corresponding authors' assignment based AAC (Álvarez-Llorente et al., 2023). The different variants obtained of the classification give results that improve those used as references in multiple scientometric fields. The variation called U1-F-0.8 seems especially promising due to its restraint in assigning multiple categories, consistency with reference classifications and the fact of applying normalization processes to avoid the overinfluence of articles that have a greater number of references.

**Financing**

Grant project PID2020-115798RB-I00 funded by Ministerio de Ciencia e Innovación of Spain (MICIN), Agencia Estatal de Investigación (AEI) / 10.13039/501100011033.

**Declarations of interest**

None.

## 1. Introduction

The assignment of papers to scientific categories is crucial for scientometrics. In this regard, the fundamental role played by large databases such as Web Of Science (WOS) and Scopus and their thematic classifications is undeniable. As is well known, both WOS and Scopus classify their papers at the journal level, that is, they assign each publication according to the discipline or disciplines where the journal they are published belongs. However, classifications based on journals are not sufficiently precise—for example for the application of normalizations in citation indicators (Althouse et al., 2009; Lancho-Barrantes et al., 2010a)—and it is necessary to have classification systems at the level of individual publications (paper-by-paper), especially when it comes to publications in journals classified as multidisciplinary or as miscellaneous with a more or less broad subject area.

Some works have highlighted inaccuracies in the categories that databases assign to journals, compared to the main topics addressed in the papers they contain (Wang & Waltman, 2016). Thelwall & Pinfield (2024) review and discuss some of these works and focus on broadly analyzing deviations in Scopus. Their findings suggest that these inaccuracies occur mainly in journals without a clearly specialised focus, particularly those categorised wholly or partly within the Multidisciplinary Area and/or the miscellaneous categories. The authors emphasise the resultant issues of search imprecision and the complications posed for reliable indicator calculations in research evaluation.

Many studies have focused their efforts on the classification of papers published in this type of journals (Glänzel et al. 1999a, 2021; Fang, 2015; Zhang et al., 2022; Zhang & Shen, 2024). Furthermore, some journals not classified as multidisciplinary should perhaps be thus classified due to the thematic variety they actually cover, as stated in the study by Zhang & Shen (2024).

This does not mean that each paper should be classified in a single thematic category and, although some studies such as those by Milojević (2020) or Waltman & Van Eck (2012) focus their efforts on this, many others (for example Fang, 2015; Glänzel et al., 2021; Zhang et al., 2022; Álvarez-Llorente et al., 2024) opt to allow for multiple assignments. Álvarez-Llorente et al. (2023) shows how the authors of the papers themselves tend to classify their publications into multiple categories. Zhang et al. (2022) also conclude that it is important to assign papers to several categories because the authors themselves do so. WOS and Scopus also allow for multiple assignments in their journal classifications.

However, it does not seem reasonable to accept the classification of papers in a high number of categories (especially if the relationship with some categories is very weak), as this could cause the imprecision of classification by journals which is precisely what we want to overcome. This is evident in some papers that limit non-obvious multiple assignments, for example Fang (2015), Glänzel et al. (2021) or Álvarez-Llorente et al. (2024).

In the scientific literature, there have been descriptions of clustering or community detection procedures used for the classification of documents registered in bibliographic databases for scientometric purposes. Mainly used to this end have been citation (Boyack & Klavans, 2010; Waltman & Van Eck, 2012; Klavans & Boyack, 2006, 2016), text (Boyack et al., 2011), or hybrid (Glenisson et al., 2005; Janssens et al., 2006, 2008, 2009; Boyack et al., 2013; Boyack & Klavans, 2020) systems.

Within citation systems, different widely accepted techniques are included and combined, such as direct citation, co-citation and bibliographic coupling (Šubelj et al., 2016, provide a discussion about these methods), applied not only to the citation relationships between publications, but also between journals (Marshakova-Shaikevich, 2005; Moya-Anegón et al., 2006; Schildt et al., 2006), keywords and/or authors (He & Hui, 2002; Zhao & Strotmann, 2022), etc., in addition to the classification of individual papers, the classification of journals (Zhang et al., 2010), specific publications such as patents (Lai & Wu, 2005), and even for the generation of thesauruses (Rees-Potter, 1989). Of all of these, direct citation is, due to its conceptual simplicity, the one that allows to manage the relationships between publications with a lighter computational complexity, which makes it easier to address larger sets of publications.

Many studies deduced that the best results are obtained through hybrid techniques. For example, Boyack & Klavans (2020) state that the combination of direct citation and textual relationship outperforms both when separate and other combinations with other citation relationships. In contrast, Chumachenko et al. (2022) states that textual relationships present problems due to the noise introduced by the habitual presence of phrases that are common to all types of disciplines, and Sachini et al. (2022), who use neural networks trained from text to reclassify Artificial Intelligence publications, indicate that these are techniques which are computationally heavy. As stated by Kandimalla et al. (2021), most unsupervised neural networks systems base their self-learning on citation networks.

The classifications obtained using automatic clustering techniques have different problems, such as the tendency to change greatly as new documents are added into the classification, and have a randomness factor that, even starting with the same sets of publications, can produce dissimilar results each time the procedure is restarted (Álvarez-Llorente et al., 2024). There has also been research on item classification using the same journal thematic classification schemes as the large WOS (Milojević, 2020) and Scopus (Álvarez-Llorente et al., 2024) bibliographic databases. In one form or another, these studies applied the categories assigned to the references used in each paper in order to classify it (in a direct citation scheme), i.e., if most of the references used in a paper are published in journals assigned to the Library and Information Science category, that paper would be considered to be from that same category.

Although this is generally true, there are cases in which it is more than a little forced. For example, techniques of Network Analysis are widely used in Information Science, while a great part of the seminal papers on Network Analysis are published in Physics journals (some of them cited in this present study) assigned to the Condensed-Matter Physics category. Does this mean that an Information Science paper citing various of those seminal Network Analysis studies could end up being categorized under Condensed-Matter Physics? This would hardly seem logical.

Instead, couldn't the category in which the references are cited be used to classify the papers rather than the category to which they belong? I.e., if in one discipline knowledge is imported from another (Guerrero-Bote et al., 2007; Lancho-Barrantes et al., 2010a) because there are references which are used a lot, that should not make papers of the importing discipline be classified under the exporting discipline, but quite the contrary since this is a behaviour typical of the importing discipline. Furthermore, it could be the case that the imported knowledge is cited more in the importing discipline than in the exporting one. This same concept is used by Li et al. (2019) to identify the disciplines in which the applications presented in software papers are used according to the publications that cite them.

A classification based on this idea would be applied in two phases: first, the references should be classified according to the categories that cite them, and second, this classification should then be used to classify the papers. This way of proceeding would have a positive side effect: all the references can be used, not just those indexed and classified in bibliographic databases. Direct citation, co-citation and bibliographic coupling systems can only take advantage of citations that are in turn indexed in the same database, since otherwise their classification would not be known.

In this present paper we shall be studying the application of a fractional method of classifying papers based on the categories of the papers that cite the references used in them, that, being consistent with the classification by journals currently established, refines and overcomes its limitations. We shall consider twelve variants in which various parameters are adjusted, and compare them with three classifications that have been presented in previous research (Álvarez-Llorente et al., 2023, 2024):

- The fractional ASJC classification, in which the Multidisciplinary Area and the miscellaneous categories have been eliminated. This is the classification most similar to the one made by Scopus, and journals, and which is currently used for scientometrics, that allows to analyze the coherence with the classifications that are currently used.
- The Author's Assignation Collection (AAC) classification done by the corresponding authors themselves. The classification that the authors themselves give to their work completely independent from the classification by journals should be the ideal classification, and any other classification should be agree with it. Unfortunately, such a classification does not currently exist for science as a whole, and is not automatable, so the AAC—a classification carried out by the authors studied in a previous research by Álvarez-Llorente et al. (2023) on a set of publications from 2020—is a valuable tool to compare other ASJC-based classifications.
- The M3-AWC-0.8 method, based on first- and second-generation references. This is a reliable classification of individual ASJC publications based on references, but with the traditional semantics of direct references, and built with parameters close to the one proposed, described in a previous research by Álvarez-Llorente et al. (2024).

### *1.1. Research Objectives*

The aim will be to respond to the following research questions, their responses aim to reveal whether the proposed classification is viable, interesting and seems capable of improving in some aspects other known previous proposals:

a) Can an algorithm be found that allows papers to be classified based on the categories of the papers that cite their references which are acceptable for computing resources and scalable to larger datasets?

b) Is it interesting to classify papers using this method? Are the classifications received with this method consistent with others or are they very different?

c) How are the classifications obtained? What is their granularity? How big are the areas and categories? Which settings lead to the most interesting results?

d) Is a classification achieved that is more consistent with the corresponding authors' AAC than other classifications?

e) Are classifications obtained with scientometrically more desirable characteristics than those based on ASJC journal assignments?

f) What proportion of papers cannot be classified? Is this proportion higher or lower than in other classifications?

g) How are the papers in the Multidisciplinary Area and the miscellaneous categories classified? Are there similarities with the other classifications? Are there area changes recorded when classifying papers from journals in the miscellaneous categories?

## 2. Method and Data

The experiment is done with scientific production from the year 2020 registered in Scopus. Specifically, we use a snapshot of the database taken in April 2022 (to which SCImago has access due to its relationship with Elsevier, owner of Scopus), from which we selected for the experiment the 3,246,022 citable papers (articles, reviews, conference papers and short surveys) indexed in the 31,185 active scientific journals from which the SJR is also calculated, including a total of 127,277,415 references to 40,283,256 different sources. The main reason for choosing this dataset is to allow for comparison with the reference classifications resulting from previous research (the Author's Assignment Collection –AAC–, described in Álvarez-Llorente et al., 2023; and the M3-AWC-0.8 method, described in Álvarez-Llorente et al., 2024), both based on Scopus publications from 2020. Furthermore, it focuses on a single year since the purpose of the classification system is, at least initially, for it to be applied yearly, such as is done, for example, to calculate the SJR rankings from Scopus or JCR from WOS. We consider that the year 2020 does not have any peculiarity regarding scientific production that could distort the results of our research, and it is a recent year that represents the current publication habits, but that at the same time allows to have a reasonably stable dataset. It is well known that in 2020 there was an explosion of publications related to COVID-19 (Farooq et al., 2021), particularly in the fields of health and life sciences initially, but also extending into many other fields as the year progressed (Aristovnik et al., 2020). This surge subsequently led to inflated impact factors of many journals. However, we consider this phenomenon to be not relevant to our study, as it focuses on comparing different classification methods within the same time window. Any imbalance should, therefore, affect all methods in a similar manner, and, consequently, the proposed method should accurately reflect the actual structure of science, whatever that may be. Impact factors are neither the source nor the focus of our study.

We shall try to classify the papers from the dataset using the ASJC scheme (discarding, as will be seen below, the Multidisciplinary Area and the miscellaneous categories).

As indicated in the previous section, the method we present differs from traditional methods based on the categories of the references (Glänzel et al., 1999a, 1999b, 2021; Milojević, 2020; Álvarez-Llorente et al., 2024) in that these categories will be established beforehand on the basis of the papers in which they are cited. Thus, the first thing the method has to do is to determine the categories to which each of the works citing the references of the papers to be classified correspond.

The starting point is the fractional ASJC journal assignment described by Álvarez-Llorente et al. (2023) which distributes the Multidisciplinary Area assignments among all the 285 categories considered and the miscellaneous of each area among all the area's categories.

According to this assignment, each journal has an associated vector of 285 elements that collects the degree of membership (weight) of the journal to each of the 285 non-miscellaneous categories, so that the sum of all the weights is 1. If a journal belongs exclusively to one category it will have a weight of 1 in the corresponding component; a journal assigned to 2 categories will have two components with a weight other than zero (greater or lesser depending on the degree of membership in the category) that add up to 1; a journal that belongs exclusively to the Multidisciplinary Area has all its cells with a value of 1/285; etc. Papers initially inherit the vector of the journal in which they have been published.

Next, for each paper we take each of its references and accumulate the weights of the paper in the vector of each reference. At the end of the process we normalize all the reference vectors so that their weights add up to 1 again.

Once all the references have been categorized in this way, for each paper we calculate its weight vector as the sum of the weights of all its references, and we also normalize it so that its weights add up to 1.

Now each paper has a new vector of weights, and, therefore, a potentially different assignment from the initial one, therefore we repeat the sequence as many times as necessary until the variations become insignificant.

Note that, although the starting point is the ASJC journal-based classification, the result of the application of this method is in an item-by-item classification in which each publication is assigned to one (or several) of the ASJC categories (not Multidisciplinary or miscellaneous) based on the references it contains, regardless of how other papers from the same journal are classified.

As this study is limited to Scopus publications from year 2020, we are able to compare the results with those of previous studies (Álvarez-Llorente et al., 2023, 2024) in which the same publication window was used. Even though this was the main objective of the limitation to this data set, it also facilitates the calculations to a degree. We also consider it to be appropriate to work with a reasonably well established snapshot which at the same time fairly faithfully reflects current citation trends.

In order to legitimize the comparisons, in coherence with the criteria that we applied in previous studies, and especially with regard to the M3-AWC-0.8 classification (Álvarez-Llorente et al., 2024), we must avoid the reclassification of publications with fewer than 3 references which would not provide a minimum of significance. Thus, 4.32% of the publications will be left unreclassified, a significantly smaller percentage than the 6.90% discarded for the M3-AWC-0.8 classification.

The following pseudocode schematizes and details the algorithm followed. In Appendix A, a more formal description of the algorithm's pseudocode can be found.

Let $P$ be the set of registered papers, where each paper $p \in P$ has a classification in the fractional ASJC (inherited from its journal), which is a 285-component vector $W_p$ with the weights in each category, normalized so that the sum of the components is 1.

$R$ is the set of references registered among the total papers of $P$, and $R_p$ the set of $N_p$ references contained in the paper $p \in P$.

1. Repeat…

    1.a. For each reference $r \in R$, a 285-component vector $\omega_r$ is generated, whose components are the sum of the weights $W_p$ of each citing paper $p$.

    1.b. We normalize each reference vector $\omega_r$ to the sum of its components (i.e., its components then sum to unity).

    1.c. For each source paper $p \in P$ with $N_p$ references $R_p$, a 285-component vector $W_p$ is generated (replacing previous $W_p$) whose components are the sum of the weights of the $N_p$ reference vectors $\omega_n$, (for $n$ from 1 to $N_p$), but only those of the categories to which the journal is assigned in the database.

    1.d. We normalize each source paper vector $W_p$ to the sum of its components (i.e., its components then sum to unit).

    …until the sum of squared differences between the new $W_p$ and older $W_p$ for every $p \in P$ is less than 3000.

2. The resulting $W_p$ vectors of step 1.d of the last iteration yields the result labeled **JL (Journal Limited)**.

3. With the foregoing result, a new 285-component vector $\omega_r$ is generated for each reference $r \in R$, whose components are the sum of the weights $W_p$ of each citing paper $p$.

4. We normalize each vector $\omega_r$ to the sum of its components (i.e., its components then sum to unit).

5. A 285-component vector $W_p$ is generated (replacing previous $W_p$) for each source paper $p \in P$ with $N$ references $R_p$, whose components are the sum of the weights of the $N$ reference vectors $\omega_n$.

6. We normalize each paper vector $W_p$ to the sum of its components (i.e., its components then sum to unit), yielding the result labeled **U1 (Unlimited 1 iteration)**.

In step 1.c., we only sum in $W_p$ the weights of the references in each of the categories where the journal has some score. The reason is because every author knows the scope of the journal by the basis of which Scopus assigns journals to the ASJC categories. Furthermore, the review process is designed for that scope -it is not logical to think that a paper on any given topic can be published in whatever journal. For this reason, in the iterative phase, membership of items to other categories is not allowed. Once the loop termination condition of step 1 is met (after a certain number of iterations), the result labeled JL (Journal Limited) is extracted. In steps 2 to 6, another iteration is carried out, but this time the items may belong to all of the categories, and the result is extracted with the label U1 (Unlimited 1 iteration). The reason for generating this second classification (U1) is to allow the proper classification of possible publications on topics other than those of the journal in which they are published.

In the initial stages of our research we were able to verify that not exercising this limitation in the first iterations produced enormously divergent classifications in each cycle. The limitation favours the convergence of the algorithm and coherence with the initial classification and, although it initially prevents the papers from moving to new categories, it does allow, within those assigned, for some to be strengthened versus others.

Thus, although, as previously discussed, multiple studies support the presence of some degree of imprecision in Scopus's category assignments, particularly concerning multidisciplinary journals (Wang & Waltman, 2016; Thelwall & Pinfield, 2024), we contend that the JL classification, while constrained to initially assigned categories but incorporating evolving weights for these assignments (and especially in the case of multidisciplinary journals), holds significant interest in bibliometric analysis. It provides a valuable reference for comparative evaluation alongside the U1 variant and the original classification from which it is derived.

Additionally, we make a "fractional" variation of this algorithm by generating in steps 1.a and 3 the vector $\omega_r$ of weights of the papers in the references by summing the weights of the papers but dividing by each citing paper's number of references $N_p$. This is done to prevent papers containing a high number of references from exerting a greater influence on the set of reference vectors compared to those with a smaller number, as this could bias the classifications towards areas where citation habits tend to use a greater number of citations (Althouse, et al., 2009; Andersen, 2023; Lancho-Barrantes, 2010b).

The loop in step 1 repeats the reclassification until the variations become insignificant. At the end of each reclassification, for each paper the sum of the squared difference of its assignment weight vector generated with respect to that of the previous iteration is calculated. We set the loop to terminate when this result is less than 3000. Table 1 presents the evolution of the squared difference in each iteration for the two versions of the algorithm – fractional and non-fractional weights. In both cases, the algorithm converges after just 6 iterations, with the squared difference starting in the first iterations above 50 000, and after 6 iterations easily crossing the threshold of 3000. We consider this value sufficient, taking into account that more than three million documents are involved in the calculation, which sums to more than fifty million degrees of non-zero membership.

| Iteration | DS weight | DS fractional weight |
|---|---|---|
| 1 | 61750.3 | 54276.6 |
| 2 | 21035.9 | 19337.2 |
| 3 | 9913.2 | 9350.7 |
| 4 | 5404.9 | 5200.7 |
| 5 | 3219.5 | 3149.0 |
| 6 | 2041.7 | 2025.1 |

*Table 1: Difference squared in running the algorithm.*

The algorithm leaves us with four classifications: those labeled JL with both the complete weight and the fractional weight, plus those generated in step 6 labeled U1 with again both the complete weight and the fractional weight. It should be borne in mind that the main difference between the JL and the U1 classifications is that, in the former, no paper has weight in any category other than that assigned to the journal, whereas in the latter, it does.

These classifications can assign a large number of categories to each paper. To limit multiple assignments, we apply the procedure described in Álvarez-Llorente et al. (2024) which was in turn based on the one used in Glänzel et al. (2021), to restrict them to a maximum of 5 categories, applying 3 distinct threshold levels, 0.5, 0.67, and 0.8, to avoid non-obvious multiple assignments. Without going into details, this mechanism means that a paper with weight in different categories is only assigned to those categories with the highest weight, as long as that weight differs sufficiently (according to the threshold) from the next category with immediately lower weight. The higher the threshold, the more it limits the possibility of accepting multiple assignments. In other words, the higher the threshold, the lower will be the average number of categories per paper in the classifications obtained.

Combining all these variants, we shall get a total of 12 different classifications.

Note that in all its variants, the calculations of each iteration are not only computationally simple and easily parallelizable (even embarrassingly parallelizable), but also the complexity of the algorithm is close to linear with respect to the number of publications covered, which greatly facilitates its scalability and the possibility of it being applied to much larger datasets.

The proposed algorithm shares several similarities with the well-known Label Propagation algorithm used for community detection in semi-supervised machine learning systems (Zhu & Ghahramani, 2002). However, there is a key distinction: the Label Propagation algorithm includes a step that processes nodes in a random order, potentially producing different results with each run. In our classification approach, by contrast, there is no element of randomness, ensuring that the same dataset will consistently yield identical results. This behaviour is clearly outlined in the pseudocode description of the algorithm provided in Appendix A.

## 3. Results and Discussion

In this section, we shall present an exhaustive study of the scientometric characteristics of the classification proposal in all its variants, comparing them with those of the selected classifications of reference. In addition to the properties of the new form of classifying, this will let us determine whether it provides a result consistent with the established and currently accepted classifications, or, on the contrary, results in a totally disparate classification.

### *3.1. Size and uniformity of the categories*

Table 2 lists some characteristics of the structure of the 12 classifications obtained, and contrasts them with the M3-AWC-0.8 and the initial fractional ASJC classification – number of non-empty categories, the size (accumulated weight) of the largest and the smallest category, the coefficient of variation of the size of each category (CV), and the granularity, understood as defined in Waltman et al. (2020) and Milojević (2020), i.e., the whole number of works divided by the sum of the squares of the weights of each category, so that the greater granularity values are related to the more balanced categories .

| Classification | Weight | Threshold | Categories | Max. Cat. | Min. Cat. | CV | Granularity |
|---|---|---|---|---|---|---|---|
| ASJC | - | - | 285 | 75837.7 | 273.2342 | 0.93 | 5.26E-05 |
| M3-AWC-0.8 | - | (0.8) | 282 | 159032.4 | 0.0592 | 1.61 | 2.72E-05 |
| JL | Non-fract. | 0.5 | 281 | 142194.0 | 0.1956 | 1.52 | 2.94E-05 |
| JL | Non-fract. | 0.67 | 281 | 146224.8 | 0.1956 | 1.56 | 2.82E-05 |
| JL | Non-fract. | 0.8 | 281 | 148347.5 | 0.1607 | 1.59 | 2.75E-05 |
| JL | Fractional | 0.5 | 281 | 141140.1 | 0.2309 | 1.51 | 2.96E-05 |
| JL | Fractional | 0.67 | 281 | 145842.6 | 0.2309 | 1.56 | 2.83E-05 |
| JL | Fractional | 0.8 | 281 | 148308.7 | 0.2309 | 1.58 | 2.76E-05 |
| U1 | Non-fract. | 0.5 | 281 | 155716.1 | 0.1812 | 1.64 | 2.63E-05 |
| U1 | Non-fract. | 0.67 | 281 | 159215.5 | 0.1812 | 1.68 | 2.54E-05 |
| U1 | Non-fract. | 0.8 | 281 | 160810.0 | 0.0845 | 1.70 | 2.49E-05 |
| U1 | Fractional | 0.5 | 281 | 154220.7 | 0.1572 | 1.63 | 2.66E-05 |
| U1 | Fractional | 0.67 | 281 | 158076.9 | 0.1572 | 1.66 | 2.57E-05 |
| U1 | Fractional | 0.8 | 281 | 160058.6 | 0.1572 | 1.69 | 2.52E-05 |

*Table 2: Structure of the classifications obtained, giving the number of non-empty categories, the accumulated weight (size) of the largest and smallest category, the coefficient of variation (CV) of the size, and the granularity.*

A first striking result of the new classifications is the fact that all cases reach the number of the same 281 non-empty ASJC categories, listed in Table 3. As can be seen, these are categories which (in 2020) have been assigned to very few journals (even none) and which, even after the division and elimination of the Multidisciplinary Area and the miscellaneous categories, continue to accumulate very little weight in the assignment of categories.

| Code | ASJC category | ASJC journals | Fractional ASJC journals | Weight fractional ASJC journals |
|---|---|---|---|---|
| 2744 | Reviews and References (medical) | 4 | 3936 | 50.09 |
| 2918 | Pathophysiology | 0 | 380 | 6.59 |
| 3502 | Dental Assisting | 1 | 358 | 26.56 |
| 3615 | Respiratory Care | 7 | 248 | 5.33 |

*Table 3. ASJC categories without weight in the new classifications, with indication of the number of journals that have the category assigned, the number of journals that have some assignment of the category after the fractioning of the Multidisciplinary Area and the miscellaneous categories, and the weight that such fractional assignments accumulate.*

In addition to this coincidence of empty categories, the new classifications show quite similar characteristics to each other as well as with respect to the M3-AWC-0.8 classification, and deviate notably from the fractional ASJC. The biggest differences among the new classifications are between the JL and the U1: the former achieve lower accumulations in the most populated category and greater in the least populated one, and lower CVs that result in slightly higher granularities, i.e., more homogeneous, and therefore more desirable, size categories. Between the fractionated and non-fractionated versions, the former show a lower CV. For the three categorization thresholds, the lowest CV is obtained with 0.5, followed by 0.67, and finally 0.8. Hence, in this first analysis, the most desirable results were obtained with the combination JL, fractional, and 0.5 threshold. It can also be observed that all JL classifications present a better CV than M3-AWC-0.8, just the opposite of U1. However, the differences between the variants do seem of little importance since they always give granularities of the same order of magnitude.

To analyse more precisely how homogeneous the classifications are, we calculated the CVs of the number of references per paper in each category (Table 4). The table differentiates the CVs for general references and those directed to Scopus, and also those limited to the previous 2 and 3 years. Unlike in the previous analysis, the lowest CVs now correspond to the U1 classifications and with higher thresholds, although also in this case the variants with fractional weight are better than those with non-fractional weight. Once again, the differences are minimal, very similar to that of the M3-AWC-0.8 classification and in almost all cases better than the fractional ASJC.

| Classification | Weight | Threshold | ACV S | ACV S3 | ACV S2 | ACV | ACV 3 | ACV 2 |
|---|---|---|---|---|---|---|---|---|
| ASJC | - | - | 0.81 | 1.12 | 1.25 | 0.70 | 1.03 | 1.16 |
| M3-AWC-0.8 | - | (0.8) | 0.80 | 1.10 | 1.25 | 0.69 | 1.01 | 1.15 |
| JL | Non-fract. | 0.5 | 0.79 | 1.12 | 1.26 | 0.69 | 1.02 | 1.16 |
| JL | Non-fract. | 0.67 | 0.79 | 1.12 | 1.27 | 0.68 | 1.02 | 1.16 |
| JL | Non-fract. | 0.8 | 0.79 | 1.11 | 1.26 | 0.68 | 1.02 | 1.17 |
| JL | Fractional | 0.5 | 0.79 | 1.11 | 1.25 | 0.68 | 1.02 | 1.16 |
| JL | Fractional | 0.67 | 0.79 | 1.11 | 1.25 | 0.68 | 1.02 | 1.16 |
| JL | Fractional | 0.8 | 0.79 | 1.11 | 1.24 | 0.68 | 1.02 | 1.16 |
| U1 | Non-fract. | 0.5 | 0.80 | 1.12 | 1.25 | 0.69 | 1.02 | 1.16 |
| U1 | Non-fract. | 0.67 | 0.79 | 1.11 | 1.26 | 0.68 | 1.02 | 1.16 |
| U1 | Non-fract. | 0.8 | 0.79 | 1.10 | 1.25 | 0.68 | 1.01 | 1.17 |
| U1 | Fractional | 0.5 | 0.79 | 1.10 | 1.24 | 0.68 | 1.01 | 1.15 |
| U1 | Fractional | 0.67 | 0.78 | 1.10 | 1.23 | 0.68 | 1.01 | 1.14 |
| U1 | Fractional | 0.8 | 0.78 | 1.09 | 1.23 | 0.67 | 1.00 | 1.14 |

*Table 4. Average coefficients of variation (ACV) of the number of references per paper in each category for each classification (S: of the number of references to the Scopus Source Set; S3: of the number of references to the Scopus Source Set in the previous 3 years; S2: of the number of references to the Scopus Source Set in the previous 2 years; V3: of the number of references in the previous 3 years; V2: of the number of references in the previous 2 years)*

We deduce from these initial analyses that the new methods provide very promising results, although we cannot choose between any of the variants. At most we can say that the fractional versions seem to give more uniform results.

### 3.2. Coincidence with the AAC

Table 5 presents the results of our analysis of the coincidence of the new methods with the corresponding authors' AAC classification (Álvarez-Llorente et al., 2023). As concluded in that study, the AAC is a classification that can be used as a reference to evaluate classification systems for scientific documents registered in Scopus that also use the ASJC scheme. To this end, the documents classified need to be restricted to the sample used in the AAC classification, which proved to be sufficiently representative.

For each method, firstly a percentage of coincidence is indicated (in the terms defined in the cited work) between the weights assigned by each classification variant and those assigned by the authors in the AAC. This percentage is complemented by two pairs of columns indicating the coincidence *from* the AAC to the new classifications and from these *towards* the AAC: what is the average rank in the classifications that the winning categories of the different AAC papers end up at (1st AAC cat. rank) and how many times is that winning category not among the 5 possible assigned in the classification (1st AAC cat. w/o C), which would indicate a null match; and, conversely, the average rank in the AAC of the winning categories for each classification (1st class. cat. rank) and how many times does that category not appear among those assigned by the AAC (1st class. cat. w/o C). Therefore, the strongest coincidences will occur with the highest values in the coincidence percentage combined with lower values in the average ranks and in the null match counts.

One can see that the agreement with the AAC is better in the new classifications than in the M3-AWC-0.8 and fractional ASJC. Broadly speaking, within the new classifications the fractioning of weights has barely any effect, but for the other variants the results are somewhat erratic:

- The coincidence percentage improves very slightly with low thresholds, and one appreciates no significant differences between the JL and U1 versions.
- The value "1st AAC cat. rank" improves considerably with high thresholds and in the JL versions.

- The value "1st AAC cat. w/o C" improves quite a bit with low thresholds and in the U1 versions.
- The values "1st class. cat. w/o C" undergo hardly any variations, with the U1 versions being slightly better but with fractional weight.
- The values "1st class. cat. rank" are almost the same, and indeed one would have to resort to a third decimal place to find variations.

Therefore, among the new classifications, the closest similarities are given by the high thresholds. It is true that the high thresholds negatively affect the number of winning categories in AAC without coincidence in the classifications, but this is logical since high thresholds lead to fewer categories assigned per paper.

| Classification | Weight | Threshold | % Coinc. | 1st AAC cat. rank | 1st AAC cat. w/o C | 1st class. cat. rank | 1st class. cat. w/o C |
|---|---|---|---|---|---|---|---|
| ASJC | - | - | 35.16 | 1.88 | 6662 | 1.19 | 118356 |
| M3-AWC-0.8 | - | (0.8) | 37.38 | 1.59 | 11539 | 1.11 | 6363 |
| JL | Non-fract. | 0.5 | 46.35 | 1.46 | 10423 | 1.11 | 5083 |
| JL | Non-fract. | 0.67 | 45.90 | 1.29 | 11574 | 1.11 | 5083 |
| JL | Non-fract. | 0.8 | 45.30 | 1.16 | 12417 | 1.11 | 5083 |
| JL | Fractional | 0.5 | 46.37 | 1.46 | 10332 | 1.11 | 5038 |
| JL | Fractional | 0.67 | 45.99 | 1.29 | 11526 | 1.11 | 5038 |
| JL | Fractional | 0.8 | 45.48 | 1.16 | 12362 | 1.11 | 5038 |
| U1 | Non-fract. | 0.5 | 46.11 | 1.62 | 9525 | 1.11 | 5073 |
| U1 | Non-fract. | 0.67 | 45.96 | 1.38 | 11125 | 1.11 | 5073 |
| U1 | Non-fract. | 0.8 | 45.26 | 1.20 | 12228 | 1.11 | 5073 |
| U1 | Fractional | 0.5 | 46.23 | 1.62 | 9439 | 1.11 | 5024 |
| U1 | Fractional | 0.67 | 46.11 | 1.38 | 11083 | 1.11 | 5024 |
| U1 | Fractional | 0.8 | 45.54 | 1.20 | 12173 | 1.11 | 5024 |

*Table 5. Comparison of classification variants with the AAC Showing the coincidence percentage, the average rank that the winning categories in the AAC obtain in the variant, the number of winning categories in the AAC that have no mention in the variant, the average rank that the winning categories have in the AAC in each variant, and the number of winning categories of the variant that are uncategorized in the AAC.*

Table 6 presents how each method behaves in terms of the number of categories per publication. For each method, the total number of assigned categories, the average number of categories assigned per paper, and the percentages of papers with 1, 2, 3, 4, and 5 or more assignments are indicated. It should be borne in mind that only the ASJC classification can present more than 5 assignments, and therefore it is the only one that can present an average of assignments greater than 5 (and it is so high because the assignments to the Multidisciplinary category have been translated into low-weight assignments to all categories, and the miscellaneous categories to all those of the related area).

| Classification | Weight | Threshold | Nº assign. | Avg. assign. | %1 | %2 | %3 | %4 | %5+ |
|---|---|---|---|---|---|---|---|---|---|
| ASJC | - | - | 248621 | 18.49 | 15.01 | 18.83 | 11.83 | 7.09 | 47.25 |
| AAC | - | - | 26141 | 1.94 | 44.85 | 30.67 | 14.41 | 5.39 | 4.68 |
| M3-AWC-0.8 | - | (0.8) | 20918 | 1.61 | 66.53 | 18.75 | 6.77 | 2.99 | 4.95 |
| JL | Non-fract. | 0.5 | 27594 | 2.11 | 52.85 | 19.80 | 7.59 | 4.19 | 15.58 |
| JL | Non-fract. | 0.67 | 21297 | 1.63 | 67.35 | 18.33 | 5.42 | 2.96 | 5.94 |
| JL | Non-fract. | 0.8 | 17518 | 1.34 | 79.56 | 13.80 | 3.03 | 1.37 | 2.25 |
| JL | Fractional | 0.5 | 28153 | 2.15 | 51.65 | 19.82 | 7.78 | 4.48 | 16.27 |
| JL | Fractional | 0.67 | 21682 | 1.66 | 66.66 | 18.11 | 5.70 | 3.01 | 6.52 |
| JL | Fractional | 0.8 | 17734 | 1.35 | 79.04 | 13.78 | 3.27 | 1.45 | 2.46 |
| U1 | Non-fract. | 0.5 | 32215 | 2.46 | 47.72 | 15.90 | 6.41 | 3.31 | 26.66 |
| U1 | Non-fract. | 0.67 | 23164 | 1.77 | 64.21 | 17.25 | 6.12 | 2.98 | 9.44 |
| U1 | Non-fract. | 0.8 | 18072 | 1.38 | 77.78 | 14.03 | 3.92 | 1.67 | 2.61 |
| U1 | Fractional | 0.5 | 32516 | 2.48 | 47.09 | 15.84 | 6.59 | 3.31 | 27.17 |
| U1 | Fractional | 0.67 | 23417 | 1.79 | 63.99 | 16.67 | 6.39 | 3.12 | 9.83 |
| U1 | Fractional | 0.8 | 18216 | 1.39 | 77.58 | 13.95 | 3.89 | 1.62 | 2.96 |

*Table 6: Comparison of the classification variants with the AAC showing the number of assignments, average number of categories per paper, and percentages of papers with 1, 2, 3, 4, and 5 or more assignments.*

Again, the classifications with a 0.8 threshold (which produce the fewest assignments per paper) present the most desirable results in terms of number of assignments, total assignments, average number of assignments, and percentage of unique assignments, improving on the M3-AWC-0.8 classification (which applies an identical threshold) and the AAC, while the ASJC presents very different values. Indeed, in view of the results obtained up to this point, we believe that these 0.8 threshold classifications are those that provided the scientometrically most desirable results, so that in the rest of the study we shall only focus on them.

### *3.3. Distribution by category and subject area*

To go more deeply into how the classifications distribute the papers among the different categories, Table 7 lists the correlations of the sum of weights in the ASJC categories between the classification systems. It can been seen that the new classifications show behaviours that are extremely similar to each other and very similar to the M3-AWV-0.8. With respect to the AAC and the fractional ASJC, the similarity is less but still quite high.

|  | AAC | ASJC | M3-AWC-0.8 | JL-NF-0.8 | JL-F-0.8 | U1-NF-0.8 | U1-F-0.8 |
|---|---|---|---|---|---|---|---|
| AAC | 1 | 0.889 | 0.821 | 0.840 | 0.850 | 0.815 | 0.829 |
| ASJC | 0.889 | 1 | 0.873 | 0.913 | 0.918 | 0.871 | 0.880 |
| M3-AWC-0.8 | 0.821 | 0.873 | 1 | 0.933 | 0.928 | 0.930 | 0.926 |
| JL-NF-0.8 | 0.840 | 0.913 | 0.933 | 1 | 0.998 | 0.991 | 0.991 |
| JL-F-0.8 | 0.850 | 0.918 | 0.928 | 0.998 | 1 | 0.988 | 0.992 |
| U1-NF-0.8 | 0.815 | 0.871 | 0.930 | 0.991 | 0.988 | 1 | 0.996 |
| U1-F-0.8 | 0.829 | 0.880 | 0.926 | 0.991 | 0.992 | 0.996 | 1 |

*Table 7: Correlations of the sum of weights in the ASJC categories between the classification systems (limited to documents included in the AAC).*

The same conclusion can be drawn from Figure 1 which shows the number of categories that accumulate different bands of weight percentage for each classification. For instance, the band "0.5" counts how many categories in each classification accumulated a weight percentage in the interval [0.4%, 0.5%). It can be clearly seen that the distribution is very similar between the 4 variants of the new classifications and the M3-AWC-0.8, while the greatest differences are found with the AAC and ASJC classifications.

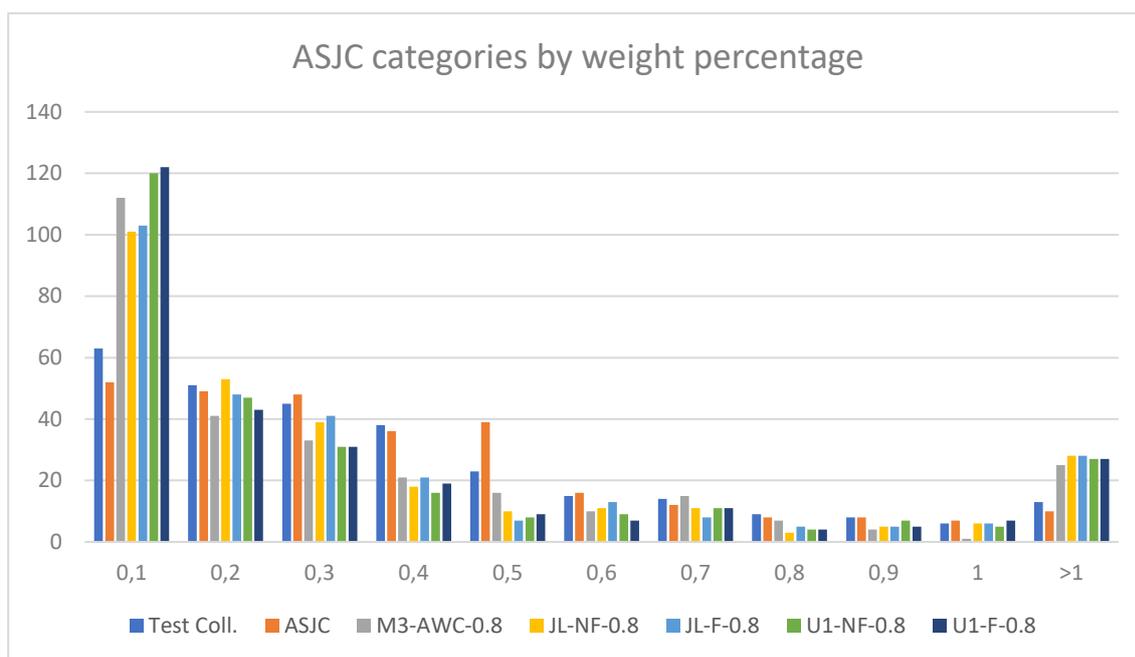

*Fig. 1. Number of categories by weight percentage for each classification method.*

Breaking down the accumulation of weights into the main areas of the ASJC (Table 8 and Figure 2), it can again be observed a great similarity among the new classifications and the M3-AWC-0.8, but less so with the two others. In some areas, such as Agricultural and Biological Sciences (1100), Medicine (2700), and Chemistry (1600), there is a visibly greater accumulation with the new classifications, to the detriment of other areas, mainly Biochemistry, Genetics and Molecular Biology (1300), and Mathematics (2600), as well as others such as Chemical Engineering (1500), Immunology and Microbiology (2400), Nursing (2900), and Health Professions (3600). Variations can also be seen between the JL and U1 versions, with the latter being those that introduce the most significant variations, both in increases and in reductions, without much difference between the fractional and non-fractional versions.

| ASJC | Description | AAC | ASJC | M3-AWC-0.8 | JL-NF-0.8 | JL-F-0.8 | U1-NF-0.8 | U1-F-0.8 |
|------|-------------|------|------|------------|-----------|----------|-----------|----------|
| 1100 | Agricultural and Biological Sciences | 5.23 | 4.90 | 7.26 | 6.30 | 6.23 | 7.23 | 7.16 |
| 1200 | Arts and Humanities | 1.21 | 1.19 | 0.79 | 0.74 | 0.75 | 0.72 | 0.72 |
| 1300 | Biochemistry, Genetics and Molecular Biology | 6.45 | 6.28 | 6.27 | 4.82 | 4.80 | 4.11 | 4.05 |
| 1400 | Business, Management and Accounting | 2.37 | 2.31 | 2.27 | 2.15 | 2.08 | 2.22 | 2.07 |
| 1500 | Chemical Engineering | 1.72 | 1.94 | 0.98 | 0.95 | 0.98 | 0.66 | 0.68 |
| 1600 | Chemistry | 3.45 | 3.88 | 3.92 | 4.91 | 4.84 | 5.17 | 5.07 |
| 1700 | Computer Science | 6.80 | 6.11 | 5.12 | 6.49 | 6.51 | 6.19 | 6.28 |
| 1800 | Decision Sciences | 0.77 | 0.57 | 0.36 | 0.29 | 0.27 | 0.25 | 0.23 |
| 1900 | Earth and Planetary Sciences | 3.14 | 3.57 | 4.20 | 3.30 | 3.30 | 3.44 | 3.41 |
| 2000 | Economics, Econometrics and Finance | 2.18 | 1.62 | 3.07 | 1.94 | 1.99 | 2.26 | 2.31 |
| 2100 | Energy | 2.24 | 1.97 | 2.20 | 1.96 | 1.91 | 2.01 | 1.87 |
| 2200 | Engineering | 7.60 | 8.43 | 8.87 | 8.75 | 8.87 | 9.06 | 9.22 |
| 2300 | Environmental Science | 3.94 | 4.45 | 2.44 | 4.00 | 3.94 | 3.50 | 3.44 |
| 2400 | Immunology and Microbiology | 2.21 | 1.42 | 1.79 | 1.30 | 1.31 | 1.28 | 1.29 |
| 2500 | Materials Science | 5.36 | 5.01 | 2.86 | 5.44 | 5.51 | 5.12 | 5.30 |
| 2600 | Mathematics | 5.60 | 4.49 | 4.04 | 3.65 | 3.71 | 3.58 | 3.66 |
| 2700 | Medicine | 17.33 | 19.78 | 21.97 | 22.10 | 22.20 | 22.91 | 23.12 |
| 2800 | Neuroscience | 1.73 | 1.57 | 1.16 | 1.18 | 1.11 | 1.10 | 1.00 |
| 2900 | Nursing | 1.44 | 0.91 | 0.56 | 0.52 | 0.52 | 0.41 | 0.42 |
| 3000 | Pharmacology, Toxicology and Pharmaceutics | 1.65 | 1.91 | 1.63 | 1.70 | 1.68 | 1.67 | 1.55 |
| 3100 | Physics and Astronomy | 4.37 | 6.43 | 6.48 | 6.36 | 6.39 | 5.70 | 5.80 |
| 3200 | Psychology | 2.51 | 2.25 | 2.41 | 1.98 | 1.97 | 1.80 | 1.78 |
| 3300 | Social Sciences | 8.99 | 7.31 | 8.13 | 7.91 | 7.91 | 8.48 | 8.42 |
| 3400 | Veterinary | 0.26 | 0.43 | 0.13 | 0.33 | 0.33 | 0.25 | 0.25 |
| 3500 | Dentistry | 0.34 | 0.45 | 0.53 | 0.48 | 0.49 | 0.49 | 0.50 |
| 3600 | Health Professions | 1.10 | 0.83 | 0.55 | 0.43 | 0.43 | 0.39 | 0.38 |

*Table 8: Percentage of accumulation (sum of weights) of the AAC documents into the 26 ASJC areas according to the classification systems compared.*

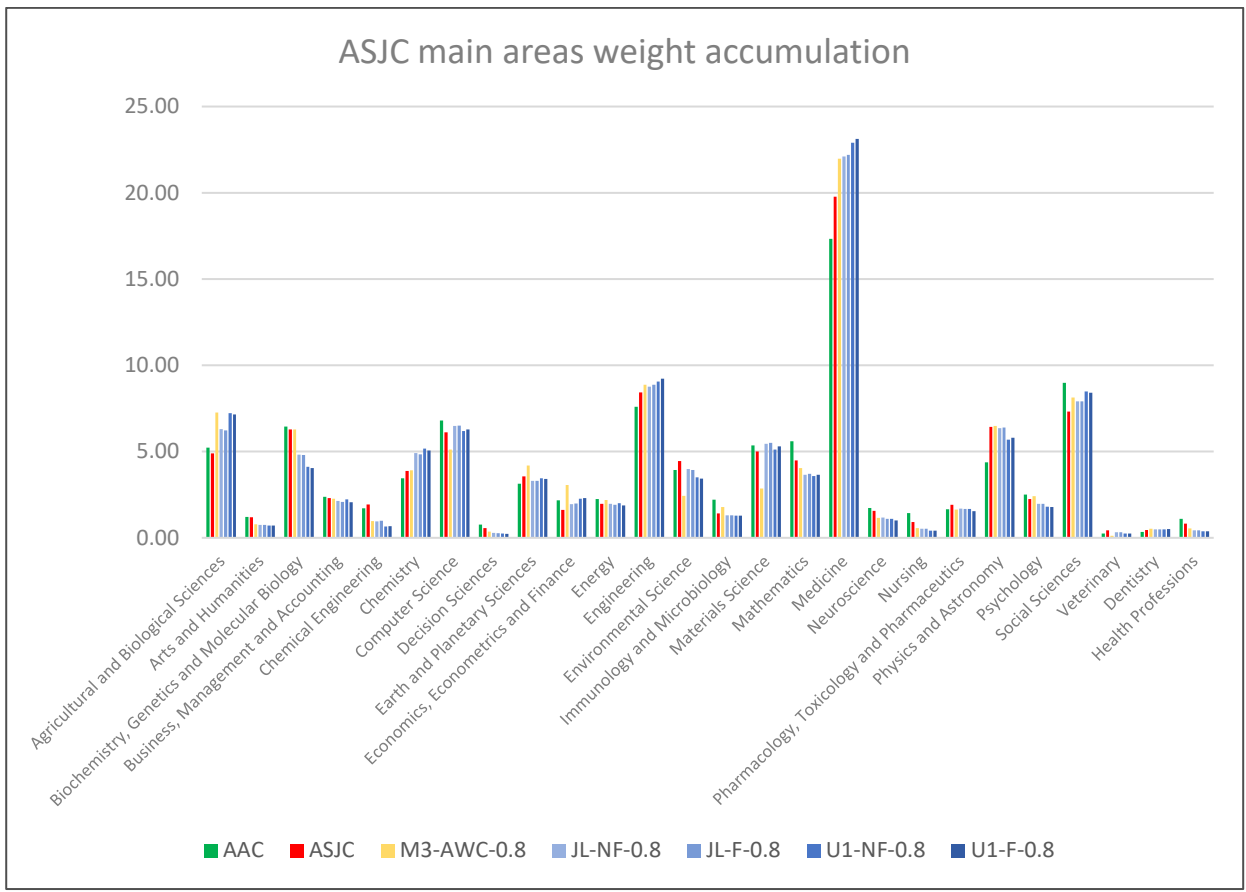

*Fig. 2. Bar chart representation of the Table 8 data in which one can appreciate the differences between the methods being compared.*

### 3.4. Relocation of Multidisciplinary Area and miscellaneous publications

With all of the above, it can be seen that the new classifications introduce no major disruption with respect to the previous ones with which they are being compared. To reinforce this idea, we shall analyse what happens with the publications most susceptible to large relocations – those that were originally assigned to the Multidisciplinary Area and each area's miscellaneous categories. Table 9 and Figure 3 show the accumulation of weights in the main areas of the ASJC after the different reclassifications of the publications initially assigned exclusively to the Multidisciplinary category. In this case, the fractional ASJC classification is not shown because in it all publications are distributed with identical weight into all categories without any more elaborate criterion.

| ASJC | Description | AAC | M3-AWC-0.8 | JL-NF-0.8 | JL-F-0.8 | U1-NF-0.8 | U1-F-0.8 |
|---|---|---|---|---|---|---|---|
| 1100 | Agricultural and Biological Sciences | 8.20 | 15.42 | 16.84 | 16.54 | 17.42 | 17.40 |
| 1200 | Arts and Humanities | 0.87 | 0.36 | 0.54 | 0.54 | 0.55 | 0.55 |
| 1300 | Biochemistry, Genetics and Molecular Biology | 14.64 | 15.42 | 10.30 | 9.27 | 7.79 | 7.51 |
| 1400 | Business, Management and Accounting | 0.39 | 0.41 | 0.94 | 0.73 | 0.90 | 0.84 |
| 1500 | Chemical Engineering | 1.31 | 0.49 | 0.33 | 0.39 | 0.18 | 0.18 |
| 1600 | Chemistry | 1.49 | 2.61 | 3.96 | 4.32 | 4.24 | 4.55 |
| 1700 | Computer Science | 4.04 | 3.17 | 3.62 | 4.20 | 2.97 | 3.54 |
| 1800 | Decision Sciences | 0.18 | 0.08 | 0.08 | 0.08 | 0.08 | 0.08 |
| 1900 | Earth and Planetary Sciences | 4.18 | 5.86 | 5.71 | 5.39 | 5.32 | 5.20 |
| 2000 | Economics, Econometrics and Finance | 3.23 | 3.11 | 2.51 | 2.69 | 2.91 | 3.06 |
| 2100 | Energy | 0.81 | 0.52 | 0.34 | 0.33 | 0.29 | 0.34 |
| 2200 | Engineering | 4.57 | 4.47 | 4.38 | 4.87 | 4.50 | 4.52 |
| 2300 | Environmental Science | 7.18 | 2.57 | 3.46 | 3.52 | 3.23 | 3.00 |
| 2400 | Immunology and Microbiology | 5.37 | 2.81 | 2.75 | 2.52 | 2.97 | 2.77 |
| 2500 | Materials Science | 3.50 | 1.97 | 3.65 | 3.49 | 3.95 | 3.64 |
| 2600 | Mathematics | 3.55 | 0.86 | 0.76 | 0.72 | 0.69 | 0.69 |
| 2700 | Medicine | 13.96 | 19.56 | 22.77 | 24.04 | 24.89 | 25.50 |
| 2800 | Neuroscience | 4.15 | 3.83 | 3.18 | 2.83 | 3.79 | 2.73 |
| 2900 | Nursing | 2.31 | 1.10 | 0.77 | 0.79 | 0.68 | 0.68 |
| 3000 | Pharmacology, Toxicology and Pharmaceutics | 1.71 | 1.61 | 0.79 | 0.58 | 0.49 | 0.54 |
| 3100 | Physics and Astronomy | 3.70 | 4.43 | 4.22 | 4.49 | 4.57 | 4.96 |
| 3200 | Psychology | 2.82 | 3.49 | 1.74 | 1.50 | 0.78 | 0.91 |
| 3300 | Social Sciences | 5.60 | 3.76 | 4.80 | 4.64 | 4.98 | 4.87 |
| 3400 | Veterinary | 0.07 | 0.08 | 0.11 | 0.14 | 0.08 | 0.08 |
| 3500 | Dentistry | 0.36 | 0.92 | 0.74 | 0.81 | 1.00 | 1.09 |
| 3600 | Health Professions | 1.80 | 1.10 | 0.70 | 0.57 | 0.77 | 0.77 |

*Table 9: Distribution of papers published in journals assigned exclusively to the 1000 Multidisciplinary category into the 26 ASJC areas, comparing the classification systems.*

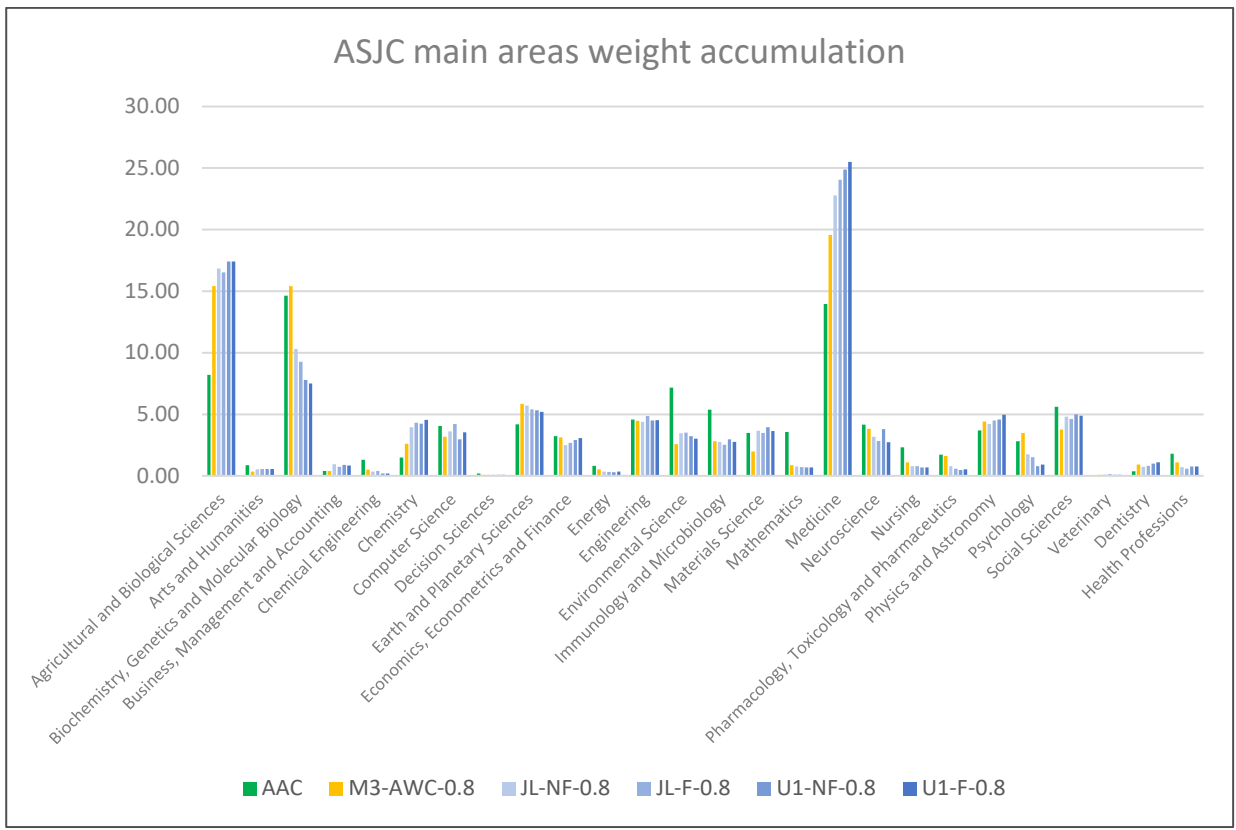

*Fig. 3. Bar chart representation of the Table 9 data highlighting the differences between the methods being compared.*

It can be observed that there is great parallelism with what was the case for the total set of AAC publications: great similarity between the new classifications and with the M3-AWC-0.8, and somewhat less so with the AAC, with greater accumulations in Agricultural and Biological Sciences (1100), Medicine (2700), and Chemistry (1600) and lesser in Biochemistry, Genetics and Molecular Biology (1300), Mathematics (2600), Chemical Engineering (1500), Immunology and Microbiology (2400), Nursing (2900), and Health Professions (3600), more significant in the U1 versions and with no clear predominance between the fractional and non-fractional classifications. A very striking decrease is also observed in Immunology and Microbiology (2400).

Extending the analysis of the redistribution to the level of the 285 categories of the fractional ASJC, we obtained the correlation data between the different classifications presented (Table 10). Again, the new classifications are very strongly related, although there is a greater gap with M3-AWC-0.8 and, above all, with the AAC. The correlations are stronger in the JL versions than in the U1.

|              | AAC   | M3-AWC-0.8 | JL-NF-0.8 | JL-F-0.8 | U1-NF-0.8 | U1-F-0.8 |
|--------------|-------|------------|-----------|----------|-----------|----------|
| **AAC**      | 1     | 0.609      | 0.563     | 0.562    | 0.517     | 0.510    |
| **M3-AWC-0.8** | 0.609 | 1        | 0.892     | 0.867    | 0.842     | 0.836    |
| **JL-NF-0.8** | 0.563 | 0.892     | 1         | 0.989    | 0.978     | 0.974    |
| **JL-F-0.8**  | 0.562 | 0.867     | 0.989     | 1        | 0.972     | 0.979    |
| **U1-NF-0.8** | 0.517 | 0.842     | 0.978     | 0.972    | 1         | 0.990    |
| **U1-F-0.8**  | 0.510 | 0.836     | 0.974     | 0.979    | 0.990     | 1        |

*Table 10: Correlations between the distributions into the 285 ASJC categories of the papers published in journals assigned exclusively to the 1000 Multidisciplinary category as done by the classification systems being compared.*

Finally, Table 11 presents the behaviour of the different classifications with respect to the papers from the miscellaneous categories. Indicated for each classification is the percentage of publications that were assigned to a category belonging to the same main area to which it belonged, which would denote consistency with the initial classification. In general, the new classifications show percentages that are very high, higher than those achieved by the AAC and much better than those of the M3-AWV-0.8. But without doubt the most striking thing is that the two versions of the JL variant (with and without fractioning) keep 100% of the publications in the same main area. This was to be expected since this case considered papers that were published in journals assigned exclusively to a miscellaneous category. Recall that the assignments to this category were divided among the rest of the categories of the same area. Hence, as this classification does not allow the assignment of a paper to a category to which the journal is not ascribed, it can only be assigned to a category in the same area.

| ASJC | Description | Nº | AAC | M3-AWC-0.8 | JL-NF-0.8 | JL-F-0.8 | U1-NF-0.8 | U1-F-0.8 |
|---|---|---|---|---|---|---|---|---|
| 1100 | Agricultural and Biological Sciences | 15 | 72.44 | 59.94 | 100.00 | 100.00 | 80.00 | 86.44 |
| 1300 | Biochemistry, Genetics and Molecular Biology | 29 | 72.47 | 77.75 | 100.00 | 100.00 | 82.81 | 82.89 |
| 1400 | Business, Management and Accounting | 9 | 72.78 | 55.56 | 100.00 | 100.00 | 80.65 | 79.91 |
| 1500 | Chemical Engineering | 14 | 65.36 | 43.36 | 100.00 | 100.00 | 53.49 | 52.45 |
| 1600 | Chemistry | 52 | 64.62 | 44.89 | 100.00 | 100.00 | 91.74 | 92.62 |
| 1700 | Computer Science | 63 | 80.61 | 49.01 | 100.00 | 100.00 | 87.85 | 91.47 |
| 1900 | Earth and Planetary Sciences | 51 | 78.73 | 72.84 | 100.00 | 100.00 | 96.69 | 93.67 |
| 2000 | Economics, Econometrics and Finance | 15 | 84.33 | 86.26 | 100.00 | 100.00 | 100.00 | 100.00 |
| 2100 | Energy | 10 | 77.50 | 60.82 | 100.00 | 100.00 | 80.00 | 80.00 |
| 2200 | Engineering | 25 | 55.47 | 48.92 | 100.00 | 100.00 | 82.20 | 82.20 |
| 2300 | Environmental Science | 15 | 53.33 | 28.34 | 100.00 | 100.00 | 78.12 | 77.96 |
| 2500 | Materials Science | 54 | 61.26 | 28.21 | 100.00 | 100.00 | 84.58 | 86.22 |
| 2600 | Mathematics | 89 | 95.00 | 87.21 | 100.00 | 100.00 | 93.62 | 94.93 |
| 2700 | Medicine | 253 | 80.37 | 86.91 | 100.00 | 100.00 | 97.06 | 97.79 |
| 2800 | Neuroscience | 32 | 76.67 | 60.15 | 100.00 | 100.00 | 90.05 | 86.47 |
| 2900 | Nursing | 16 | 78.12 | 31.25 | 100.00 | 100.00 | 51.69 | 56.20 |
| 3000 | Pharmacology, Toxicology and Pharmaceutics | 11 | 58.18 | 33.08 | 100.00 | 100.00 | 81.15 | 90.91 |
| 3100 | Physics and Astronomy | 280 | 37.98 | 46.62 | 100.00 | 100.00 | 74.57 | 79.34 |
| 3200 | Psychology | 37 | 84.50 | 66.71 | 100.00 | 100.00 | 83.18 | 82.13 |
| 3300 | Social Sciences | 12 | 82.08 | 74.55 | 100.00 | 100.00 | 91.67 | 91.67 |
| 3400 | Veterinary | 20 | 59.00 | 22.19 | 100.00 | 100.00 | 91.16 | 94.10 |
| 3500 | Dentistry | 28 | 44.23 | 64.88 | 100.00 | 100.00 | 85.28 | 86.98 |
| | Total | 1130 | 66.41 | 61.67 | 100.00 | 100.00 | 85.99 | 87.80 |

*Table 11: Percentage of the AAC papers published in journals assigned exclusively to the miscellaneous categories that the classification variants assigned to categories belonging to the same ASJC area. (Nº counts the number of papers.)*

Throughout almost all the analyses carried out, and especially in the last ones, the JL versions gave results that are closer to the reference classifications, especially to the fractional ASJC and the M3-AWC-0.8. This is logical since the classification algorithm does not allow categories to be assigned that do not appear among the journal's initial categories. And it is precisely this blocking which means that 100% of the papers in the miscellaneous categories remain in the same areas (as seen in Table 11) and which makes the behaviour of version U1 seem to us far more interesting. And between the two versions of the 0.8 threshold U1 classification, which yield very similar results, we opt for the one that takes fractional weights into account, since in our opinion the fractioning cushions the known inequalities between the different fields of science (Althouse, et al., 2009; Andersen, 2023; Lancho-Barrantes, 2010b).

In sum therefore, of the 12 variants of the proposed algorithm, we select U1-F-0.8 as the most interesting and promising.

### *3.5. Flow between categories*

The foregoing analyses have shown a high degree of similarity of the new classifications with the classifications of reference (fractional ASJC, M3-AWC-0.8, and somewhat less so with AAC). But they also showed some differences. Indeed, we selected the U1-F-0.8 variant precisely because it differentiates itself in some ambits. For this reason, we find it interesting to study what are the differences with respect to the fractional ASJC classification, the closest to the classification currently accepted in Scopus.

Figure 4 shows how the weights flowed among the ASJC areas from the original fractional ASJC classification to the reclassification using the selected U1-F-0.8 variant. The areas have been organized and coloured with the SCImago Graphica tool (Hassan-Montero et al., 2022) after grouping them into communities using the Clauset community identification algorithm (Clauset et al., 2004) and LinLog (Noack, 2007) to generate the layout, an algorithm that uses an energy model which generates layouts that are very coherent with the communities identified (Noack, 2009). The size of the circular nodes is proportional to the weight accumulated in each area in the U1-F-0.8 classification, and the thickness of the arrows is proportional to the magnitude of weight transferred from other areas in the initial fractional ASJC classification. The communities identified by the algorithm are quite natural in most cases (cluster 1 for Natural Sciences, cluster 2 for Economics, Social Sciences and Humanities, cluster 3 for Pure Sciences, Mathematics and Engineering, and cluster 4 for Health Sciences), except for leaving 2 very small clusters (cluster 5 for Psychology and cluster 6 for Dentistry) detached from others that could be very close (cluster 2 and cluster 4, respectively).

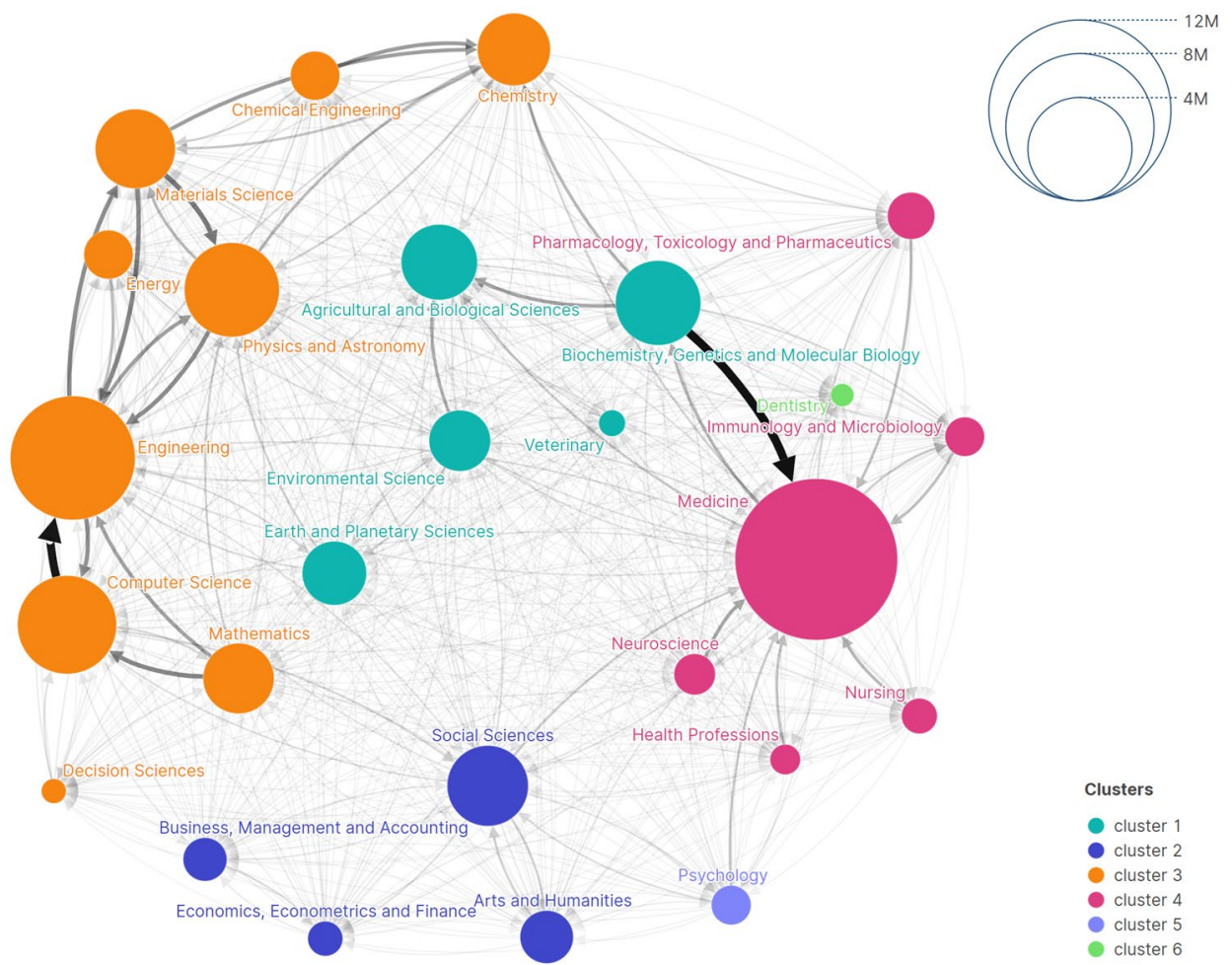

*Fig. 4. Flow of weights between the ASJC areas after the U1-F-0.8 reclassification.*

The most striking flows can be seen from Biochemistry, Genetics and Molecular Biology towards Medicine and from Computer Science towards Engineering. Also striking is the high traffic between different areas grouped within cluster 3. These flows denote a high relationship of traffic of knowledge (from one classification to the other) between the different branches of Pure Sciences, Mathematics, and Engineering.

## 4. Conclusions

The classification system for individual Scopus publications proposed in this study, based on references, but after previously categorizing the references according to the papers in which they are cited, has been shown to be practicable in terms of computation, capable of classifying up to 100% of the publications, consistent with various reference classifications, with the ability to assign multiple categories in a controlled manner and without the ambiguity of the Multidisciplinary Area and the miscellaneous categories, with good granularity in the distribution by areas, and including normalization processes so that all papers have the same weight regardless of the number of references they contain, thus in the process avoiding imbalances caused by differences in citation habits between disciplines. For all these reasons, it seems to be a viable and interesting alternative method for classifying individual papers.

Scientometrically, it has shown very interesting characteristics in terms of granularity and coherence with respect to other classification systems such as the original journal-based Scopus ASJC, the M3-AWC-0.8 based on 2-generation references, and the AAC based on the assignment made by the corresponding authors, which it improves on in many aspects. In addition, it allows for the classification of a greater number of papers by not requiring that references be directed to publications indexed in the database itself. Thus, only 4.32% of the publications were not reclassified, and this percentage could well be lower if we relaxed the self-imposed condition of excluding publications with fewer than 3 references because to us they do not seem sufficiently significant.

Despite its being based on an iterative process of refinement, the classification is not excessively costly to generate since the algorithm converges in just 6 iterations.

Twelve variants of the system were proposed, adjusting the freedom to assign categories not initially included, varying the way the weight of the assignments to categories is counted, and adjusting a threshold to limit the number of multiple assignments (up to a maximum of 5). They all provided results that were extremely similar to each other. They present high granularity, a relatively low average number of categories per paper, and cover practically all the categories of the ASJC structure, but eliminate the Multidisciplinary Area and the miscellaneous categories which are often a source of inconveniences scientometrically. And all of this is while maintaining a distribution of publications by thematic areas that is quite consistent with the original Scopus and the other reference classifications used.

Among these variants, we opted for the U1 version with fractional weight and 0.8 threshold, which we have denoted in the study as U1-F-0.8. Although it is true that the U1 variants are the furthest from the reference classifications used for comparison, they are also the ones that are most interesting from our point of view precisely because they admit a certain divergence with respect to the traditional journal-based or reference-based classifications which is especially visible in the assignment of multidisciplinary publications (where the alternative JL version is excessively rigid), without allowing any assignment outside the area.

The choice of the fractional weight versions is justified not so much because of the results obtained, which in general are not very differentiating, but mainly because of the concept itself which, from a theoretical point of view, intimates at a sensation of normalization so appreciated in scientometrics.

The 0.8 threshold (the highest of those tested) seems to us the most interesting as it provides on average fewer assignments per publication. Although in several analyses the results closest to the reference classifications were those with the lowest threshold (0.5), other analyses have shown just the contrary. From our point of view, the assignment of a reduced number of categories per paper is a very desirable feature.

Our joint analysis of the new classification system (with all its variants) in contrast to the original ASJC classification, the M3-AWC-0.8 and the AAC (all based on quite different ideas: journal classification, double-reference generation, manual assignment by the authors), has revealed a notable distance between the AAC and the others, which on the other hand turn out to be very similar classifications to each other. This reinforces the idea expressed in our and other researchers' previous studies (Álvarez-Llorente et al., 2023; Shu et al., 2019; Zhang et al., 2022) about the striking deviation between the way the corresponding authors categorize their works, the categorization of the journals in which they publish them, and the categories of the publications they cite, to which we now add the categories of other publications that use the same citations.

Finally, although the study has been limited to Scopus publications from the year 2020 with the aim of being able to compare the result with that of previous research (Álvarez-Llorente et al., 2023, 2024), we do not believe that there is any reason that prevents the classification method being applied to any other time window, as well as to any other database of publications classified by journal. Furthermore, the classification algorithm is based on computationally simple calculations, has almost linear complexity, and is easily parallelizable, which is a guarantee of scalability to much larger datasets. However, the proposed classification system is initially intended to be applied to annual windows (as has been done in the present study) in which it is likely that quantities of publications extraordinarily greater than those included here are not covered.

All of this leads to the thought that the classification method could be postulated as the basis for a classification system for the large databases that currently classify by journal.

### Appendix A. Algorithm's pseudocode

Considerations for the pseudocode:

1. **Optimization potential**: The pseudocode presented is open to various optimizations. To enhance readability, several searches are presented as simple linear searches, though these could be further optimized.
2. **Identifiers**: Each paper or reference has assigned a unique identifier.
3. **Total papers**: NumPapers represents the total number of papers.
4. **Paper weights (WPapers)**: WPapers[1..NumPapers][1..285] denotes the weights of each paper across the 285 categories. These weights are normalized so that the sum of the 285 components for each paper is 1.
5. **Number of references per paper (NumPaperRefs)**: NumPaperRefs[1..NumPapers] indicates the number of references each paper contains.
6. **Paper references (PaperRefs)**: PaperRefs[1..NumPapers][1..NumRefs[$p$]] holds the identifiers for each of the NumRefs[$p$] references of each paper $p$.
7. **Total unique references**: NumRefs represents the total count of distinct references.
8. **Reference identifiers (Refs)**: Refs[1..NumRefs] lists the unique identifiers for each reference.
9. **Reference weights (WRefs)**: WRefs[1..NumRefs][1..285] will be the vector of weights for each unique reference across the 285 categories.

```
// Step 1
REPEAT
  Wprev[][] = WPapers[][]   // NumPapers × 285 matrix copy

  // Step 1a
  FOR nr IN 1 TO NumRefs DO
    WRefs[nr][] = {0,0,…}   // 285 items vector initialization to 0
    FOR np IN 1 TO NumPapers DO
      FOR npr IN 1 TO NumPaperRefs[np] DO // This loop is a linear search
        IF Refs[nr] = PaperRefs[np][npr] THEN
          WRefs[nr][] += WPapers[np][]   // 285 items vector addition

  // Step 1b
  FOR nr IN 1 TO NumRefs DO
    Normalize(WRefs[nr][])   // 285 items normalization

  // Step 1c
```

```
  FOR np IN 1 TO NumPapers DO
    WPapers [np][] = {0,0,…}   // 285 items vector initialization
    FOR npr IN 1 TO NumPaperRefs[np] DO
      FOR nr IN 1 TO NumRefs DO   // This loop is a linear search
        IF Refs[nr] = PaperRefs[np][npr] THEN
          FOR c IN 1 TO 285 DO   // Vector addition but only where previous indexes were not 0
            IF Wprev[np][c] > 0 THEN
              WPapers[np][c] += WRefs[nr][c]
          EXIT FOR nr   // Search end

  // Step 1d
  FOR np IN 1 TO NumPapers DO
    Normalize(WPapers[np][])   // 285 items normalization

UNTIL SquareDiference(WPapers[][], Wprev[][]) < 3000

// Step 2: Extract JL classification
JL[][] = WPapers[][]   // NumPapers × 285 items matrix copy

// Step 3
FOR nr IN 1 TO NumRefs DO
  WRefs[nr][] = {0,0,…}   // 285 items vector initialization to 0
  FOR np IN 1 TO NumPapers DO
    FOR npr IN 1 TO NumPaperRefs[np] DO   // This loop is a linear search
      IF Refs[nr] = PaperRefs[np][npr] THEN
        WRefs[nr][] += WPapers[np][]   // 285 items vector addition
      EXIT FOR nr   // Search inside this paper end

// Step 4
FOR nr IN 1 TO NumRefs DO
  Normalize(WRefs[nr][])   // 285 items normalization

// Step 5
FOR np IN 1 TO NumPapers DO
  WPapers [np][] = {0,0,…}   // 285 items vector initialization
  FOR npr IN 1 TO NumPaperRefs[np] DO
    FOR nr IN 1 TO NumRefs   // This loop is a linear search
      IF Refs[nr] = PaperRefs[np][npr] THEN
        WPapers [np][] += WRefs[nr][]   // 285 items vector addition

// Step 6 and Extract U1 classification
FOR np IN 1 TO NumPapers DO
  Normalize WPapers[np][]   // 285 items normalization
U1[][] = WPapers[][]   // NumPapers × 285 items matrix copy
```

Where Normalize(vector) is:

```
Sum = 0
FOR i IN 1 TO 285 DO Sum += vector[i]
FOR i IN 1 TO 285 DO vector[i] /= Sum
```

# 5. References

Althouse, B. M., West, J. D., Bergstrom, T. C., & Bergstrom, C. T. (2009). Differences in impact factor across fields and over time. *Journal of the Association for Information Science and Technology, 60*(1), 27–34. https://doi.org/10.1002/asi.20936

Álvarez-Llorente, J. M., Guerrero-Bote, V. P., & De Moya Anegón, F. (2023). Creating a collection of publications categorized by their research guarantors into the Scopus ASJC scheme. *Profesional de la Información, 32*(7), Article e320704. https://doi.org/10.3145/epi.2023.dic.04

Álvarez-Llorente, J. M., Guerrero-Bote, V. P., & De Moya-Anegón, F. (2024). New fractional classifications of papers based on two generations of references and on the ASJC scopus scheme. *Scientometrics*. https://doi.org/10.1007/s11192-024-05030-2

Andersen, J. P. (2023). Field-level differences in paper and author characteristics across all fields of science in Web of Science, 2000-2020. *Quantitative Science Studies, 4*(2), 394–422. https://doi.org/10.1162/qss_a_00246

Aristovnik A., Ravšelj D., & Umek L. (2020). A Bibliometric Analysis of COVID-19 across Science and Social Science Research Landscape. *Sustainability, 12*(21):9132. DOI: http://dx.doi.org/10.3390/su12219132

Boyack, K. W., & Klavans, R. (2010). Co-citation analysis, bibliographic coupling, and direct citation: Which citation approach represents the research front most accurately? *Journal of the Association for Information Science and Technology, 61*(12), 2389–2404. https://doi.org/10.1002/asi.21419

Boyack, K. W., & Klavans, R. (2020). A comparison of large-scale science models based on textual, direct citation and hybrid relatedness. *Quantitative Science Studies (1)4,* 1570–1585. https://doi.org/10.1162/qss_a_00085

Boyack, K. W., Newman, D., Duhon, R. J., Klavans, R., Patek, M., Biberstine, J. R., Schijvenaars, B., Skupin, A., Ma, N., & Börner, K. (2011). Clustering more than two million biomedical publications: Comparing the accuracies of nine text-based similarity approaches. *PLOS ONE, 6*(3), Article e18029. https://doi.org/10.1371/journal.pone.0018029

Boyack, K. W., Small, H., & Klavans, R. (2013). Improving the Accuracy of Co-citation Clustering Using Full Text. *J Am Soc Inf Sci Tec, 64*: 1759–1767. https://doi.org/10.1002/asi.22896

Chumachenko, A., Kreminskyi, B., Mosenkis, I., & Yakimenko, A. (2022). Dynamical entropic analysis of scientific concepts. *Journal of Information Science, 48*(4), 561–569. https://doi.org/10.1177/0165551520972034

Clauset, A., Newman, M., & Moore, C. (2004). Finding community structure in very large networks. *Physical Review E, 70*(6). https://doi.org/10.1103/physreve.70.066111

Fang, H. (2015). Classifying Research Articles in Multidisciplinary Sciences Journals into Subject Categories. *Knowledge Organization, 42*(3), 139–153. https://doi.org/10.5771/0943-7444-2015-3-139

Farooq R. K., Rehman S. U., Ashiq M., Siddique N., & Ahmad S. (2021). Bibliometric analysis of coronavirus disease (COVID-19) literature published in Web of Science 2019-2020. *J Family Community Med., 28*(1):1-7. DOI: https://doi.org/10.4103/jfcm.JFCM_332_20

Glänzel, W., Schubert, A., & Czerwon, H. (1999a). An item-by-item subject classification of papers published in multidisciplinary and general journals using reference analysis. *Scientometrics, 44*(3), 427–439. https://doi.org/10.1007/bf02458488

Glänzel, W., Schubert, A., Schoepflin, U., & Czerwon, H. (1999b). An item-by-item subject classification of papers published in journals covered by the SSCI database using reference analysis. *Scientometrics, 46*(3), 431–441. https://doi.org/10.1007/BF02459602

Glänzel, W., Thijs, B., & Huang, Y. (2021). Improving the precision of subject assignment for disparity measurement in studies of interdisciplinary research. In: W. Glänzel, S. Heeffer, PS. Chi, R. Rousseau, *Proceedings of the 18th International Conference of the International Society of Scientometrics and Informetrics (ISSI 2021)*, Leuven University Press, 453–464. https://kuleuven.limo.libis.be/discovery/fulldisplay?docid=lirias3394551&context=SearchWebhook&vid=32KUL_KUL:Lirias&search_scope=lirias_profile&tab=LIRIAS&adaptor=SearchWebhook&lang=en

Glenisson, P., Glänzel, W., Janssens, F., De Moor, B. (2005). Combining full text and bibliometric information in mapping scientific disciplines. *Information Processing & Management, 41*(6), 1548–1572. https://doi.org/10.1016/j.ipm.2005.03.021

Guerrero-Bote, V. P., Zapico-Alonso, F., Espinosa-Calvo, M. E., Gómez-Crisóstomo, R., & De Moya-Anegón, F. (2007). Import-export of knowledge between scientific subject categories: The iceberg hypothesis. *Scientometrics, 71*(3), 423–441. https://doi.org/10.1007/s11192-007-1682-3

Hassan-Montero, Y., De Moya-Anegón, F., & Guerrero-Bote, V. P. (2022). SCImago Graphica: a new tool for exploring and visually communicating data. *Profesional de la información, 31*(5), Article e310502. https://doi.org/10.3145/epi.2022.sep.02

He, Y., & Hui, S. C. (2002). Mining a Web Citation Database for author co-citation analysis. *Information Processing & Management, 38*(4), 491–508. https://doi.org/10.1016/s0306-4573(01)00046-2

Janssens, F., Glänzel, W., & De Moor, B. (2008). A hybrid mapping of information science. *Scientometrics, 75*(3), 607–631. https://doi.org/10.1007/s11192-007-2002-7

Janssens F., Leta, J., Glänzel, W., & De Moor, B. (2006). Towards mapping library and information science. *Information Processing & Management, 42*(6), 1614–1642. https://doi.org/10.1016/j.ipm.2006.03.025

Janssens, F., Zhang, L., De Moor, B., & Glänzel, W. (2009). Hybrid clustering for validation and improvement of subject-classification schemes. *Information Processing & Management, 45*(6), 683–702. https://doi.org/10.1016/j.ipm.2009.06.003

Kandimalla, B., Rohatgi, S., Wu, J., & Giles, C. L. (2021). Large scale subject category classification of scholarly papers with deep attentive neural networks. *Frontiers in Research Metrics and Analytics, 5*, Article 600382. https://doi.org/10.3389/frma.2020.600382

Klavans, R., & Boyack, K. W. (2006). Quantitative evaluation of large maps of science. *Scientometrics, 68*, 475–499. https://doi.org/10.1007/s11192-006-0125-x

Klavans, R., & Boyack, K. W. (2016). Which Type of Citation Analysis Generates the Most Accurate Taxonomy of Scientific and Technical Knowledge? *Journal of the Association for Information Science and Technology, 68*(4), 984–998. https://doi.org/10.1002/asi.23734 .

Li, K., Chen, P.-Y., & Fang, Z. (2019). Disciplinarity of software papers: A preliminary analysis. *Proceedings of the Association for Information Science and Technology (56)*, 706–708. https://doi.org/10.1002/pra2.143

Lancho-Barrantes, B. S., Guerrero-Bote, V. P., & De Moya-Anegón, F. (2010a). The iceberg hypothesis revisited. *Scientometrics, 85*(2), 443–461. http://dx.doi.org/10.1007/s11192-010-0209-5

Lancho-Barrantes, B. S., Guerrero-Bote, V. P., & De Moya-Anegón, F. (2010b). What lies behind the averages and significance of citation indicators in different disciplines? *Journal of Information Science, 36*(3), 371–382. https://doi.org/10.1177/0165551510366077

Lai, K., & Wu, S. (2005). Using the patent co-citation approach to establish a new patent classification system. *Information Processing & Management, 41*(2), 313–330. https://doi.org/10.1016/j.ipm.2003.11.004

Milojević, S. (2020). Practical method to reclassify Web of Science articles into unique subject categories and broad disciplines. *Quantitative science studies, 1*(1), 183–206. https://doi.org/10.1162/qss_a_00014

Marshakova-Shaikevich, I. (2005). Bibliometric maps of field of science. *Information Processing & Management, 41*(6), 1534–1547. https://doi.org/10.1016/j.ipm.2005.03.027

Moya-Anegón, F., Herrero-Solana, V., & Jiménez-Contreras, E. (2006). A connectionist and multivariate approach to science maps: the SOM, clustering and MDS applied to library and information science research. *Journal of Information Science, 32*(1), 63–77. https://doi.org/10.1177/0165551506059226

Noack, A. (2007). Energy models for graph clustering. *Journal of Graph Algorithms and Applications, 11*(2), 453–480. https://doi.org/10.7155/jgaa.00154

Noack, A. (2009). Modularity clustering is force-directed layout. *Physical Review E, 79*(2). https://doi.org/10.1103/physreve.79.026102

Rees-Potter, L. K. (1989). Dynamic thesaural systems: A bibliometric study of terminological and conceptual change in sociology and economics with application to the design of dynamic thesaural systems. *Information Processing & Management, 25*(6), 677–689. https://doi.org/10.1016/0306-4573(89)90101-5

Sachini, E., Sioumalas-Christodoulou, K., Christopoulos, S., & Karampekios, N. (2022) AI for AI: Using AI methods for classifying AI science documents. *Quantitative Science Studies, 3*(4), 1119–1132. https://doi.org/10.1162/qss_a_00223

Schildt, H., & Mattsson, J. (2006). A dense network sub-grouping algorithm for co-citation analysis and its implementation in the software tool Sitkis. *Scientometrics, 67*, 143–163. https://doi.org/10.1007/s11192-006-0054-8

Shu, F., Julien, C., Zhang, L., Qiu, J., Zhang, J., & Larivière, V. (2019). Comparing journal and paper level classifications of science. *Journal of Informetrics, 13*(1), 202–225. https://doi.org/10.1016/j.joi.2018.12.005

Šubelj, L., Van Eck, N. J., & Waltman, L. (2016). Clustering Scientific Publications Based on Citation Relations: A Systematic Comparison of Different Methods. *PLOS ONE, 11*(4), Article e0154404. https://doi.org/10.1371/journal.pone.0154404

Thelwall, M., Pinfield, S. (2024). The accuracy of field classifications for journals in Scopus. *Scientometrics, 129*(2), 1097–1117. https://doi.org/10.1007/s11192-023-04901-4

Waltman, L., & Van Eck, N. J. (2012). A new methodology for constructing a publication-level classification system of science. *Journal of the Association for Information Science and Technology, 63*(12), 2378–2392. https://doi.org/10.1002/asi.22748

Waltman, L., Boyack, K. W., Colavizza, G., & Van Eck, N. J. (2020). A principled methodology for comparing relatedness measures for clustering publications. *Quantitative Science Studies, 1*(2), 691–713. https://doi.org/10.1162/qss_a_00035

Wang, Q., & Waltman, L. (2016). Large-scale analysis of the accuracy of the journal classification systems of Web of Science and Scopus. *Journal of Informetrics, 10*(2), 347-364. https://doi.org/10.1016/j.joi.2016.02.003

Zhang, L., Janssens, F., Liang, L., & Glänzel W. (2010). Journal cross-citation analysis for validation and improvement of journal-based subject classification in bibliometric research. *Scientometrics, 82*, 687–706. https://doi.org/10.1007/s11192-010-0180-1

Zhang, L., Sun, B., Shu, F., & Huang, Y. (2022). Comparing paper level classifications across different methods and systems: an investigation of Nature publications. *Scientometrics, 127*(12), 7633–7651. https://doi.org/10.1007/s11192-022-04352-3

Zhang, J., & Shen Z. (2024). Analyzing journal category assignment using a paper-level classification system: multidisciplinary sciences journals. *Scientometrics*. https://doi.org/10.1007/s11192-023-04913-0

Zhao, D., & Strotmann, A. (2022). Intellectual structure of information science 2011-2020: an author co-citation analysis. *Journal of Documentation, 78*(3), 728–744. https://doi.org/10.1108/JD-06-2021-0119

Zhu, X. & Ghahramani, Z. (2002). Learning from Labeled and Unlabeled Data with label Propagation. *CMU Technical Report CMU-CALD-02-10*.